\documentclass[fleqn,usenatbib]{mnras}

\usepackage[T1]{fontenc}

\DeclareRobustCommand{\VAN}[3]{#2}
\let\VANthebibliography\thebibliography
\def\thebibliography{\DeclareRobustCommand{\VAN}[3]{##3}\VANthebibliography}

\usepackage{graphicx}	
\usepackage{amsmath}	

\title[Optimised complexity population models with RJMCMC.]{Is there an excess of black holes around $20 M_{\odot}$? Optimising the complexity of population models with the use of reversible jump MCMC. }

\author[A. Toubiana et al.]{
A. Toubiana,$^{1}$\thanks{E-mail: atoubiana@aei.mpg.de}
Michael L. Katz,$^{1}$
Jonathan R. Gair$^{1}$
\\
$^{1}$Max Planck Institute for Gravitationsphysik (Albert Einstein Institute), Am M\"{u}hlenberg 1, 14476 Potsdam, Germany\\
}

\date{Accepted XXX. Received YYY; in original form ZZZ}

\pubyear{2023}

\begin{document}
\label{firstpage}
\pagerange{\pageref{firstpage}--\pageref{lastpage}}
\maketitle

\begin{abstract}
Some analyses of the third gravitational wave catalogue released by the LIGO-Virgo-KAGRA collaboration (LVK) suggest an excess of black holes around $15-20 M_{\odot}$. In order to investigate this feature, we introduce two flexible population models, a semi-parametric one and a non-parametric one. Both make use of reversible jump Markov chain Monte-Carlo to optimise their complexity. We also illustrate how the latter can be used to efficiently perform model selection. Our parametric model broadly agrees with the fiducial analysis of the LVK, but finds a peak of events at slightly larger masses. Our non-parametric model shows this same displacement. Moreover, it also suggests the existence of an excess of black holes around $20 M_{\odot}$. We assess the robustness of this prediction by performing mock injections and running simplified hierarchical analyses on those (i.e. without selection effects and observational uncertainties). We estimate that such a feature might be due to statistical fluctuations, given the small number of events observed so far, with a $5\%$ probability. We estimate that with a few hundreds of observations, as expected for O4, our non-parametric model will be able to robustly determine the presence of this excess. It will then allow for an efficient agnostic inference of the properties of black holes. 
\end{abstract}




\section{Introduction}

The total number of gravitational wave (GW) observations from compact binaries by the LIGO-Virgo-KAGRA (LVK) collaboration \citet{LIGOScientific:2014pky,VIRGO:2014yos,KAGRA:2020tym} now adds up to 90 \citet{LIGOScientific:2021djp}. Together they form the third GW transients catalogue (GWTC-3) released by the LVK. As the number of observation increases, we become able to infer the astrophysical properties of GW sources not only individually but as a population. Different astrophysical models for the formation and evolution mechanisms of GW sources predict different distributions of parameters such as masses, spins or redshifts \citet{Benacquista:2011kv,Postnov:2014tza,deMink:2016vkw,Samsing:2017rat,Gerosa:2018wbw,Tagawa:2019osr,Sedda:2023big,Giacobbo:2018etu,Wiktorowicz:2019dil,vanSon:2022myr,Romero-Shaw:2020siz,Mapelli:2021taw,Bouffanais:2021wcr,Inayoshi:2017mrs}. Thus, inferring the population of GW sources from data is a powerful tool to constrain astrophysical models.

Focusing on binary black holes (BBHs), one of the main results of population analyses is the growing evidence for an excess of black holes (BHs) around $35M_{\odot}$ \citet{LIGOScientific:2020kqk,KAGRA:2021duu,Tiwari:2020vym,Tiwari:2021yvr,Tiwari:2023xff,Edelman:2022ydv,Farah:2023vsc,Sadiq:2021fin,Ruhe:2022ddi,Callister:2023tgi}. This excess can be interpreted as a "pile-up" of BHs before the upper mass gap \citet{Woosley:2016hmi,Woosley:2002zz,Farmer:2019jed,Talbot:2018cva}, although it happens at lower masses than predicted by current astrophysical models \citet{Belczynski:2016jno,Marchant:2018kun,Renzo:2020lwl,Farag:2022jcc}. Moreover, first hints of correlations between parameters have started to appear \citet{Hoy:2021rfv,Callister:2021fpo,Fishbach:2021yvy,Adamcewicz:2022hce,Bavera:2022mef,Biscoveanu:2022qac}, e.g., between the effective spin and the mass ratio or between redshift and spins. Such correlations carry the signature of the astrophysical channels through which the binaries form. However, some of these were not anticipated by astrophysical models, illustrating once more how population analysis can shed light on astrophysics.

 Another interesting feature is a possible excess of BHs around $15-20 M_{\odot}$ \citet{KAGRA:2021duu,Edelman:2022ydv,Tiwari:2021yvr,Tiwari:2023xff}. It has been pointed out that it could be the signature of second generation mergers \citet{Tiwari:2020otp,Tiwari:2021yvr,Mahapatra:2022ngs}. However, it is statistically less significant than the feature at around $35 M_{\odot}$ \citet{Farah:2023vsc,Tiwari:2023xff} and has been found by only a few of the analyses performed on GWTC-3. Assessing the significance of this excess was the first motivation for our study.
 
From a Bayesian perspective, the goal of population inference \citet{Mandel:2018mve,Vitale:2020aaz} is to obtain the posterior distribution on population hyperparameters (denoted $\Lambda$) assuming a population prior for the individual event parameters $p(\Theta|\Lambda)$. The event parameters of interest, $\Theta$, are typically the masses, the spins and the redshift of the source. Broadly speaking, we can identify three main approaches, differing in how $p(\Theta|\Lambda)$ is modelled:
\begin{itemize}
    \item astrophysical: the distribution of parameters is obtained from astrophysical simulations. These simulations typically output samples from the population prior (i.e., events). 
    \item parametric: the population prior is written as a combination of simple functions, which depend on hyperparameters that are inferred from the data. 
    \item non-parametric: a more complicated and flexible functional form for the population prior is assumed, with a variable number of degrees of freedom. The parameters of the model have (in general) no physical meaning.
\end{itemize}
In the first approach, the hyperparameters have a clear astrophysical meaning. They are related to the parameters of the astrophysical simulations, for example the efficiency of energy transfer from the binary to the gas during the common envelope stage, or the branching fraction between binaries formed in isolation and those formed dynamically, or parameters characterising properties of primordial BHs \citet{Zevin:2020gbd,Wong:2020ise,Franciolini:2021tla,Mould:2022ccw}. Moreover, such an approach naturally incorporates correlations between parameters. However, it heavily relies on assumptions about astrophysical processes that are highly uncertain and has limited flexibility. Furthermore, in the standard way of performing population inference one needs to evaluate $p(\Theta|\Lambda)$ \citet{Mandel:2018mve,Vitale:2020aaz}, thus requiring an additional step where the probability density function (pdf) is estimated from samples of the population prior with the use of neural networks or kernel density estimators \citet{Zevin:2020gbd,Wong:2020ise,Franciolini:2021tla,Mould:2022ccw,Toubiana:2021iuw}. The parametric approach \citet{Talbot:2018cva,LIGOScientific:2020kqk,KAGRA:2021duu} partially leverages these issues, using functional forms for the population prior that are generic enough and could describe a variety of astrophysical scenarios. This is the approach used in the fiducial POWER-LAW+PEAK (PP) analysis of the LVK, which describes the primary mass pdf as the weighted sum of a power-law and a Gaussian \citet{KAGRA:2021duu}. Although more flexible than the astrophysical one, the parametric approach also makes assumptions about the general form of the population prior. For instance, the fiducial LVK analysis does not allow for excesses of BHs at both $20M_{\odot}$ and $35M_{\odot}$. Such assumptions are no longer required in the non-parametric approach. The price to pay is an increase in the number of parameters to be inferred and the loss of physical interpretation of those parameters. 

However, it is precisely non-parametric models that have first indicated a potential excess of BHs around $15-20 M_{\odot}$ in the LVK analysis. Additional analyses since then have also found evidence for such an excess \citet{Edelman:2022ydv,Tiwari:2021yvr,Tiwari:2023xff}, but not all \citet{Sadiq:2021fin,Ruhe:2022ddi,Callister:2023tgi}. A generic problem of non-parametric models is that the number of parameters is allowed to be arbitrarily large. Adding parameters generally increases the complexity of the model but might lead to "overfitting" the data. Therefore, a compromise needs to be found, which requires iterating over the model dimensionality to find a suitable balance between the quality of the fit and the size of the model. In this work, we propose reversible jump Markov chain Monte-Carlo (RJMCMC) as a tool to optimally determine the complexity of the model. See \citet{Rinaldi:2021bhm} for a different approach using Dirichlet processes. We consider two models for the distribution of primary masses that take advantage of RJMCMC. A semi-parametric one that is a more flexible version of the PP model of the LVK, and a non-parametric one, representing the pdf of primary masses as a piece-wise power-law function. Moreover, we illustrate how RJMCMC can be used to perform model selection without having to perform multiple runs.

Applying both models to GWTC-3, our semi-parametric model favours having a second Gaussian at $\sim 10 M{\odot}$ relative to the PP model, leading to a displacement of the peak of low-mass events. Our non-parametric model shows more differences, in broad agreement with other non-parametric analyses of the LVK. In particular, it agrees on the displacement of the peak at low masses and does show some evidence for an excess of BHs around $20 M_{\odot}$. We investigate the statistical significance of this excess under simplifying assumptions by performing mock injections, and estimate that there is a $5\%$ chance that this excess is due to statistical fluctuations, when assuming a population compatible with the PP model. We show that with $\sim 500 $ events our non-parametric model will be able to reliably and efficiently identify such an excess. 

This paper is organised as follows. In Sec.~\ref{sec:setup}, we describe the details of our analysis. We present the results of our population analysis on GWTC-3 in Sec.~\ref{sec:hba_gwtc3}. Then, in Sec.~\ref{sec:mock_hba} we describe our mock injections and comment on the significance of the excess. Finally, in Sec.~\ref{sec:ccl} we present our general conclusions.


\section{Setup}\label{sec:setup}

In this section, we will start by reviewing the basics of hierarchical Bayesian analyses, then we comment on the advantages of RJMCMC in population analyses and finally describe the population models we use. 

\subsection{Hierarchical Bayesian framework}\label{sec:hba}

Assuming a population prior $p(\Theta|\Lambda)$, we can write the number density of events as:
\begin{equation}
    \frac{{\rm d} N}{{\rm d}\Theta}  (\Lambda) = N(\Lambda) p(\Theta|\Lambda), 
\end{equation}
such that $N(\Lambda)$ is the total number of events during the observation period $T_{\rm obs}$ predicted by the population model. Inference on $\Lambda$ is performed within a hierarchical Bayesian framework. Given a set of $N_{obs}$ observed data, $\{d_i\}$, the posterior on the hyperparameters governing the population is \citet{Mandel:2018mve,Vitale:2020aaz}:
\begin{equation}
    p(\Lambda|\{d_i\}) \propto \pi(\Lambda) e^{-\xi(\Lambda) N(\Lambda)}\prod_{i=1}^{N_{obs}} \int  \frac{{\rm d} N}{{\rm d}\Theta}  (\Lambda) \frac{p(\Theta|d_i) }{\pi_{PE}(\Theta) }  {\rm d} \Theta ,
\end{equation}
where $p(\Theta|d_i)$ is the single event posterior, $\pi_{PE}(\Theta)$ is the prior used for parameter estimation, $\pi(\Lambda)$ is the prior on the hyperparameters and $\xi(\Lambda)$ is the selection function. The latter measures the fraction of events from the population that we expect to observe for a given value of $\Lambda$:
\begin{equation}
    \xi(\Lambda)=\int  \int_{d \ {\rm detectable}} p(d|\Theta) p(\Theta|\Lambda)  {\rm d} d {\rm d} \Theta .
\end{equation}
The integral over $d$ is performed only over the detectable datasets. 
It defines the probability of detecting such an event from the population characterised by the hyperparameters $\Lambda$. A handy way to compute the selection function for a variety of population models is to first perform an injection campaign, generating mock events with realistic noise and running the detection pipelines used during operational runs to determine if they are detectable. The selection function is then computed via Monte-Carlo integration:
\begin{equation}
    \xi(\Lambda)=\frac{1}{N_{all}}\sum_{\Theta_i \ {\rm detected}} \frac{p(\Theta_i|\Lambda)}{\pi_{\emptyset}(\Theta_i)},
\end{equation}
where $\pi_{\emptyset}(\Theta)$ is the prior used to perform the injections, $N_{all}$ is the \emph{total} number of injection performed and the sum runs only over the parameters that lead to detectable events. We marginalise over the statistical uncertainty in $\xi(\Lambda)$ coming from the Monte-Carlo estimation following the method in \citet{Farr:2019rap}. We compute the selection function using the injection campaign performed by the LVK \citet{data_gwtc3,data_gwtc3_pop} and applying the same criteria described in appendix A of \citet{KAGRA:2021duu} for the detectability of an event. Finally, we evaluate the posterior on $\Lambda$ via Monte-Carlo integration: 
\begin{equation}
    p(\Lambda|\{d_i\}) \propto  \pi(\Lambda) e^{-\xi(\Lambda) N(\Lambda)} \prod_i^{N_{obs}} \sum_{\Theta_j \sim p(\Theta|d_i)} \frac{1}{\pi_{PE}(\Theta_j)} \left . \frac{{\rm d} N}{{\rm d}\Theta} (\Lambda)  \right |_j  , \label{eq:pop_posterior}
\end{equation}
where it is assumed that we use the same number of samples for each event. We retain the same 69 BBH events with false alarm rate below $0.25yr^{-1}$ and use the same parameter estimation samples for those as \citet{KAGRA:2021duu}. These are provided in the public data release of the Gravitational Wave Open Science Center \citet{data_gwtc3,data_gwtc3_pop}. Finally, we apply the criteria on the number of effective samples described in appendix B of \citet{KAGRA:2021duu}. 

In Eq.~\ref{eq:pop_posterior}, we use the proportionality symbol instead of the equality one because we have omitted numerical factors that depend on the observed data $\{d_i\}$, but not on $\Lambda$, i.e., the individual event evidences and the overall model evidence. These factors are unimportant when the goal is to obtain the posterior distribution on $\Lambda$, but the overall model evidence is usually required to perform model selection. Assuming equal a priori probability for the models we want to compare, the ratio between evidences gives the Bayes' factor between models. As we will describe next, RJMCMC allows us to bypass evidence computation and returns the Bayes' factor without any extra cost. 

\begin{table*}
  \begin{center}
   \begin{tabular}{c *{8}{c|}}
   
   \hline
   
   \multicolumn{1}{|c|}{Parameter} &  \multicolumn{1}{|c|}{$\lambda$} &  \multicolumn{1}{|c|}{$\mu$} & \multicolumn{1}{|c|}{$\sigma$} &  \multicolumn{1}{|c|}{$m_{min}$}  &  \multicolumn{1}{|c|}{$m_{max}$}  &  \multicolumn{1}{|c|}{$m_{break}$}  &  \multicolumn{1}{|c|}{$\alpha$}  &  \multicolumn{1}{|c|}{$\delta_m$} \\

    \hline 
    
   \multicolumn{1}{|c|}{Range} & \multicolumn{1}{|c|}{$0-10^5$} &\multicolumn{1}{|c|}{2-100 $M_{\odot}$} & \multicolumn{1}{|c|}{1-10 $M_{\odot}$} & \multicolumn{1}{|c|}{2-10 $M_{\odot}$} & \multicolumn{1}{|c|}{30-100 $M_{\odot}$}& \multicolumn{1}{|c|}{2-100 $M_{\odot}$} &\multicolumn{1}{|c|}{1.1-10} &\multicolumn{1}{|c|}{0.5-10} \\
    
    \hline
  
   \end{tabular}
   \end{center}
    \caption{Range of the priors on the hyperparameters of the FLEXIBLE POWER-LAW+GAUSSIANS model. We assume uniform priors in the provided ranges. }\label{tab:priors_pl_gauss}
  \end{table*}

\subsection{Reversible jump MCMC}

RJMCMC \citet{Green:1995mxx} is a powerful method that explores the parameter space while allowing its dimensionality to vary. This is achieved by proposing not only changes in the parameters of the current model, but also the addition or removal of model parameters. As an example, imagine we have a datastream containing an unknown number of signals of known shape. RJMCMC allows the number of sources in the datastream to be estimated while also estimating their parameters. It returns a posterior distribution on the number of sources present in the datastream. This is needed when the data stream from a detector may contain many signals simultaneously, such as LISA and the "global fit" problem \citet{Littenberg:2023xpl,Littenberg:2020bxy,Katz2023}. RJMCMC can also be used to perform model selection: in this example, the ratio of the number of samples containing $n_1$ sources to the number of samples containing $n_2$ sources gives the ratio between the evidences of the two hypotheses. In this work we use the Eryn implementation of RJMCMC \citet{Karnesis:2023ras,eryn}. For more details on the mathematics and methods used in RJMCMC within Eryn, we refer the reader to \citet{Karnesis:2023ras}. We have also implemented our code for GPUs to speed up the computation. The results presented in this work take in average 4 days to run on a GPU.

In the context of population analysis, RJMCMC is useful for many purposes. First, it allows us to explore more flexible combinations to describe the population prior. For instance, we can consider an extended PP model where the number of Gaussians is free to vary. We can also let the presence of a power-law component be decided by the data, allowing for 0 or 1 (or even more) power-laws while sampling. Finally, we can perform model selection between having a broken power-law or a simple power-law, letting the number of each component be 0 or 1 and jumping between them. 
For non-parametric models, the number of parameters required is related to the complexity of the pdf being fitted. Increasing the number of parameters is tempting but might lead to ``overfitting'' of the data and finding spurious features and of course leads to a more complicated posterior to be sampled. Moreover, one usually has to try different configurations, performing runs for different numbers of parameters until the best compromise is found, based on the evidence. RJMCMC alleviates this burden by letting the number of parameters be a free parameter and providing a posterior over it. In this sense, the complexity of the model is chosen by the  data.

 \begin{figure*}
\centering
 \includegraphics[width=\textwidth]{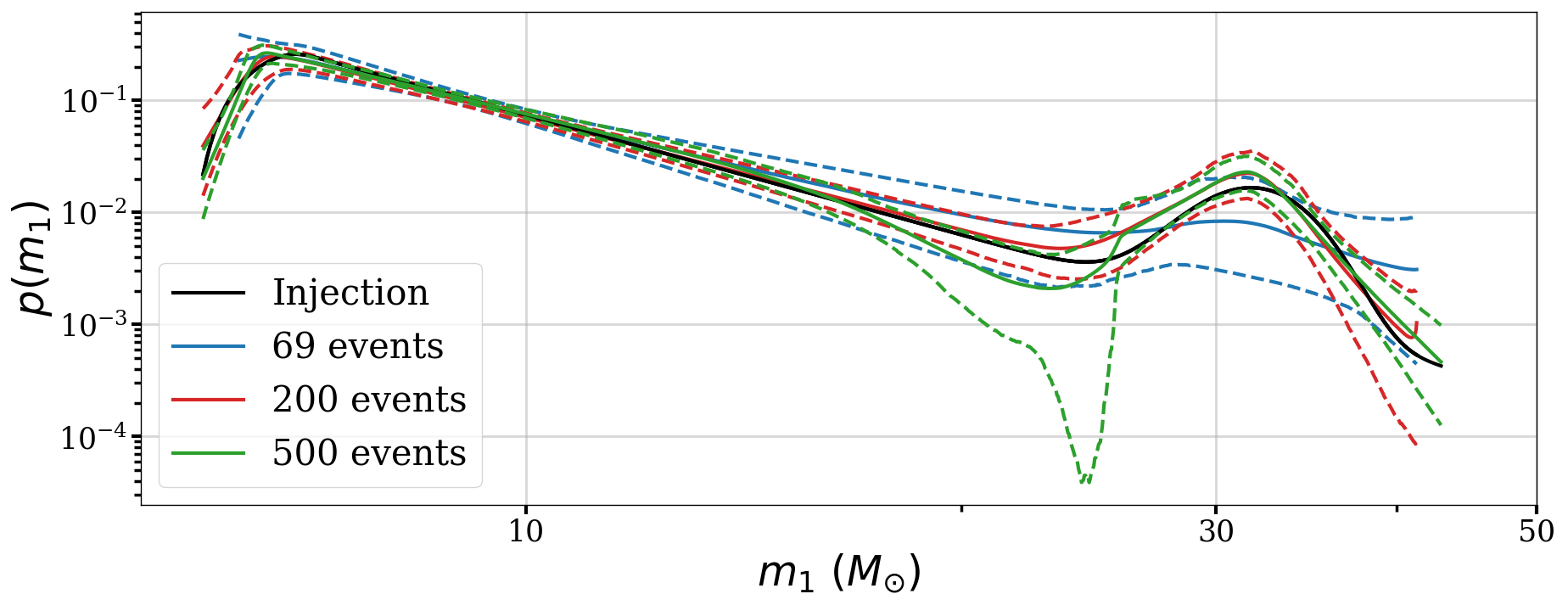}\\
 \centering
 \caption{Comparison between the $m_1$ pdf from which we draw events (black line) and the inferred distribution with our non-parametric model. The solid coloured lines show the mean and the dashed coloured lines delimit the $90\%$ confidence interval for different numbers of injections. The injected pdf is typically within the $90\%$ confidence interval. The injection pdf is given by the PP model with $\lambda^G=5.4$, $\lambda^{PL}=63.6$, $\mu=31.8 M_{\odot}$, $\sigma=2.8 M_{\odot}$, $m_{min}=3.6 M_{\odot}$, $m_{max}=99.4 M_{\odot}$, $\alpha=3.5$, $\delta_m=1.7 M_{\odot}$.  }\label{fig:mock_lvc_ppd}
\end{figure*}

 \begin{figure}
\centering
 \includegraphics[width=0.5\textwidth]{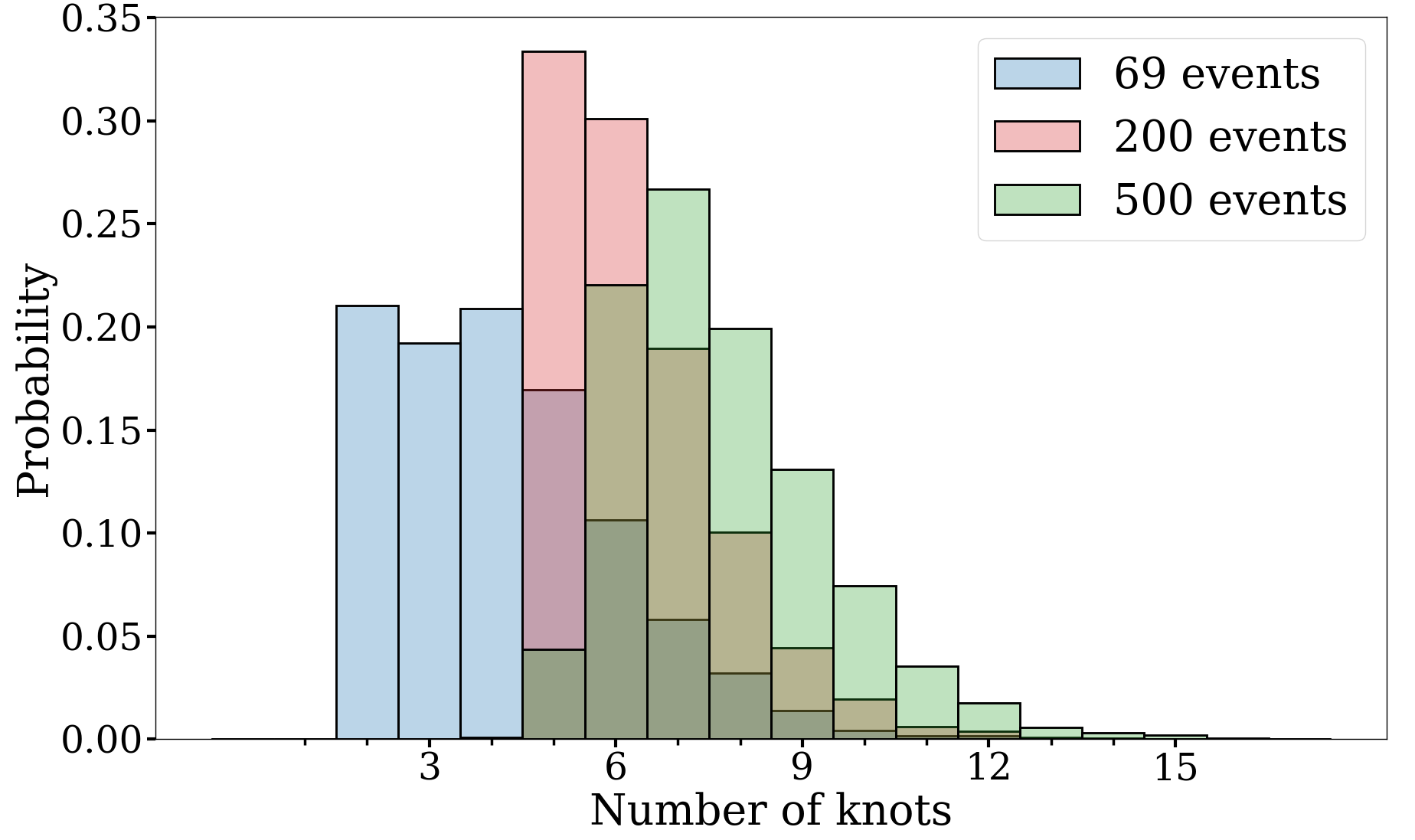}\\
 \centering
 \caption{Posterior distribution on the number of knots for each of the curves shown in Fig.~\ref{fig:mock_lvc_ppd}. Increasing the number of events barely increases the number of knots required to represent the pdf.}\label{fig:mock_lvc_hist_knots}
\end{figure}

\subsection{Population models}

As in \citet{KAGRA:2021duu}, the event parameters used to perform the population inference are the source-frame mass of the primary $m_1$, the mass ratio $q\leq1$, the spin magnitudes $\chi_1$ and $\chi_2$, the angles between the BH spins and the orbital angular momentum of the binary (tilt angles), $\theta_1$ and $\theta_2$, and the redshift of the source, $z$. We assume the number density to be separable:
\begin{align}
\frac{{\rm d} N}{{\rm d}\Theta} (\Lambda) =& \frac{{\rm d} N}{{\rm d}m_1} (\Lambda_{m_{1}})p(q|m_1,\Lambda_q,\Lambda_{m_1})p(\chi_1|\Lambda_{\chi})p(\chi_2|\Lambda_{\chi}) \nonumber \\ p&(\theta_1,\theta_2|\Lambda_{\theta})p(z|\Lambda_{z}).
\end{align}
and focus on modelling the primary mass number density, $\frac{{\rm d} N}{{\rm d}m_1} (\Lambda_{m_{1}})$. We describe next the different possibilities that we explore for it. As for the remaining parameters (mass ratio, spins and redshift) we use the same model as the fiducial analysis in \citet{KAGRA:2021duu}, which we describe in App.~\ref{app:pop_priors}. 

\subsubsection{Semi-parametric model}

We consider an extended version of the PP model of the LVK, with two main differences: we let the number of Gaussians vary, and, instead of requiring the power-law component to be present, we allow for either one power-law, or one broken power-law, or none. We label this model FLEXIBLE POWER-LAW+GAUSSIANS (FPG). The $m_1$ number density can be written:
\begin{align}
    &\frac{{\rm d} N}{{\rm d}m_1} (\Lambda_{m_{1}})=S(m_1,m_{min},\delta_m) \  \times  \nonumber \\
    & \left [ \sum_{i=0}^{N_G} \lambda_i^G G(m_1,\mu_i,\sigma_i) + \lambda^{PL} \sum_{i=0}^1 PL(m_1,m_{min},m_{max},\alpha)  \right . \nonumber \\ 
    & + \left . \lambda^{BPL} \sum_{i=0}^1 BPL(m_1,m_{min},m_{max},m_{break},\alpha_1,\alpha_2) \right ] \label{eq:ext_pl_gauss}  
\end{align}
where $N_G$ is the maximum number of Gaussians allowed a priori, $G(m,\mu,\sigma)$ is a Gaussian distribution centred around $\mu$ of width $\sigma$, $PL$ is a power-law and $BPL$ is a broken power-law, defined in Eqs.~\ref{eq:pl} and \ref{eq:bpl} below. $S(m,m_{\min},\delta_m)$ is the smoothing function introduced in \citet{Talbot:2018cva}, which we give explicitly in App.~\ref{app:pop_priors}, and $\delta_m$ defines a scale over which the $m_1$ pdf goes smoothly to zero. The power-law and broken power-law are defined as:
\begin{align}
&PL(m_1,m_{min},m_{max},\alpha) =
	\begin{cases}
		& \mathcal{N} m_1^{-\alpha} , \  {\rm if} \;  m_{min} \leq m_1 \leq m_{max}  ;  \\ 
		& 0 \ {\rm otherwise} ,
	\end{cases} \label{eq:pl} \\
&BPL(m_1,m_{min},m_{max},m_{break},\alpha_1,\alpha_2) =
	\begin{cases}
		& \mathcal{N} m_1^{-\alpha_1} \left( 1+\frac{m_1}{m_{break}} \right)^{\alpha_1-\alpha_2} \\
  &{\rm if} \;  m_{min} \leq m_1 \leq m_{max}  ;  \\ 
		& 0 \ {\rm otherwise} , \label{eq:bpl}
	\end{cases}
\end{align} 
where in both cases $\mathcal{N}$ is the appropriate normalisation factor. Note that in the case $\alpha_1=\alpha_2$ we recover the simple power-law.
In Eq.~\ref{eq:ext_pl_gauss}, it is understood that the $i=0$ case corresponds to the absence of the component. For instance, we can have two Gaussians with or without a power-law, or just a broken power-law and so on and so forth. The only restrictions are that we must have at least one component and that we cannot have the power-law and the broken power-law simultaneously. In any of these cases we set the likelihood to zero.  

In Eq.~\ref{eq:ext_pl_gauss}, the amplitudes $\lambda$s are related to the number of events in each component. Strictly speaking there is not equality because of the smoothing function $S(m,m_{\min},\delta_m)$, but this is a small correction. To avoid cases where a component is added with very small amplitude, we use a flat prior on the $\lambda$'s. We also use flat priors for the remaining hyperparameters characterising the $m_1$ distribution, with ranges given in Table~\ref{tab:priors_pl_gauss}. The prior on the number of Gaussians, power-law and broken-power-law is also taken to be flat. For all other population hyperparameters we use the same priors as in \citet{KAGRA:2021duu}.

\begin{figure*}
\centering
 \includegraphics[width=\textwidth]{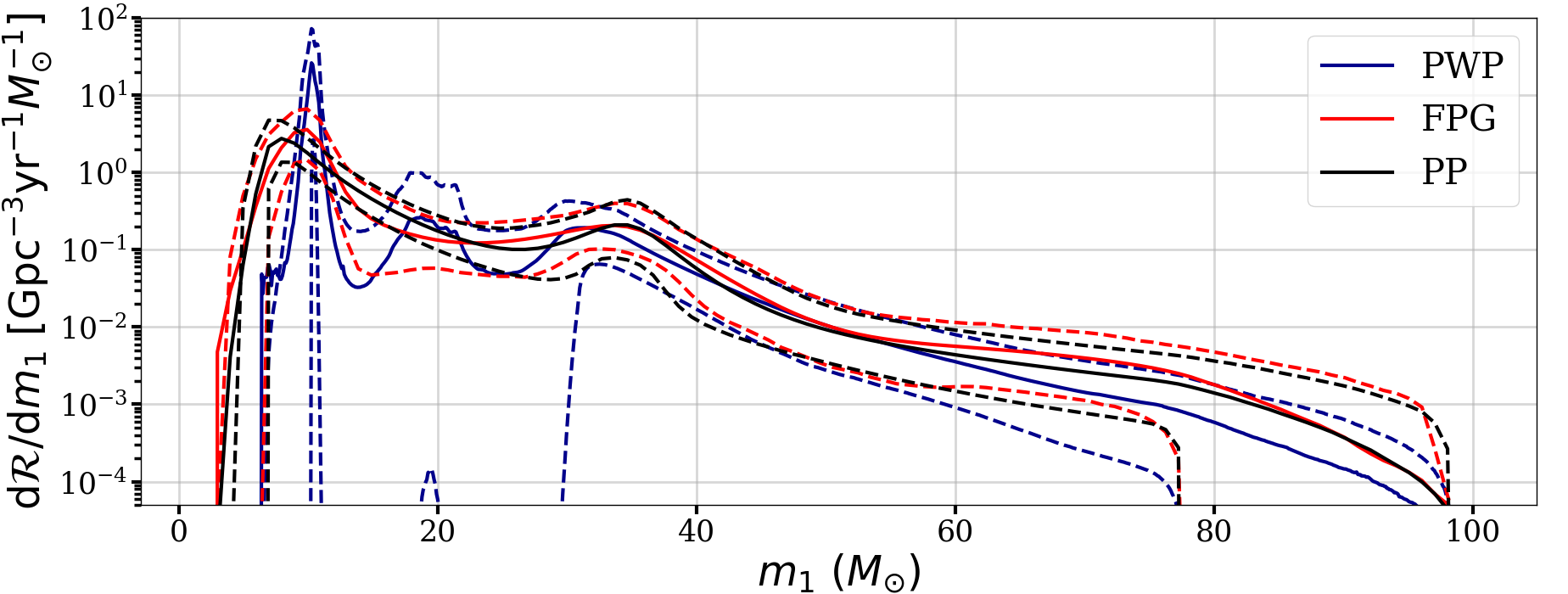}\\
 \centering
 \caption{Comparison of the volumetric rate as a function of $m_1$ predicted by the PP (in black), the FPG (in orange) and the PWP (in blue) models. Full lines indicate the mean and dashed lines the $90\%$ confidence interval of each model. The FPG model is in good agreement with the PP one, except for a displacement of the peak towards slightly larger masses. This comes from favouring a second Gaussian at $\sim 10 M_{\odot}$ in addition to the one at $\sim 35M_{\odot}$, as shown in Fig.~\ref{fig:mock_lvc_ppd}. As for the PWP model, it shows more deviations from the PP model. In particular, it agrees with the FPG model concerning the shift of the peak, favouring an even more pronounced peak, and suggests an excess of BHs around $20M_{\odot}$, as also suggested by some previous analyses~\citet{KAGRA:2021duu,Edelman:2022ydv,Tiwari:2021yvr,Tiwari:2023xff}.}\label{fig:hba_lvk}
\end{figure*}

\subsubsection{Non-parametric models}

In our non-parametric model, we describe the $m_1$ number density as a PIECE-WISE POWER-LAW (PWP) function. We write $\Lambda_{m_1}=\{x_i,v_i\}_n$, where $v_i$ is the value of the pdf at a knot $x_i$, and $n$ is the total number of knots. The number density at any point is obtained by interpolation:
\begin{equation}
 \frac{{\rm d} N}{{\rm d}\Theta} (\Lambda) =
	\begin{cases}
		& v_i\left( \frac{m_1}{x_i}\right ) ^{\frac{\log(v_{i+1}/v_i)}{\log(x_{i+1}/x_i)}} ,  \\ 
  & {\rm if} \;  x_1<...<x_i \leq m_1<x_{i+1}<...<x_n  ; \\
		& 0 \ {\rm if} \; m_1<x_1 \; {\rm or} \  m_1>x_n.
	\end{cases} \label{eq:fid_prior}
\end{equation}
This is equivalent to assuming that $\log \left(  \frac{{\rm d} N}{{\rm d}\Theta} (\Lambda) \right )$ is a piece-wise linear function of $\log(m_1)$.

We assume a log-flat prior on the $\{v_i\}_n$ and on the $\{x_i\}_n$ and a flat prior on the number of knots. In principle, the range of the prior on the position of the knots is determined by the minimum and maximum $m_1$ sample over all events. However, because there are very few samples above $100 M_{\odot}$ (less than $0.5 \%$ of the total samples) and those are sparsely distributed, we find that letting the knots take values above $100 M_{\odot}$ leads to spurious features. In fact, above $100 M_{\odot}$ the determination of the population posterior is almost completely driven by the selection function. Therefore, we take $100 M_{\odot}$ as the upper limit for the position of the knots. 

We illustrate our method by generating mock injections compatible with the PP population inferred by the LVK, and recovering the pdf with our non-parametric model. We consider an increasing number of events: 69, as in the current dataset, 200 and 500. The latter two define a realistic range for the number of BBHs we expect to have observed after the fourth operational run (O4). For simplicity, in this illustrative case we do not include either selection effects or measurement errors. In Fig.~\ref{fig:mock_lvc_ppd}, we compare the recovered pdfs to the pdf from which the mock injections are drawn. The injected pdf typically lies within the $90\%$ confidence interval, showing that our method is able to properly infer it. Moreover, increasing the number of events does not lead to a dramatic increase in the number of knots used for interpolation, as can be seen in Fig.~\ref{fig:mock_lvc_hist_knots}. Thus, the number of free parameters of our model remains reasonable as the size of the dataset increases. We have performed such injections for 200 sets of hyperparameters drawn from the LVK posterior and produced pp-plots by computing to which quantile of the recovered distribution do the quantiles of the observed set of events correspond. We obtain diagonal pp-plots, reinforcing our confidence that our non-parametric model can be used to infer LVK-like populations.  

One of the non-parametric models used by the LVK is the POWER-LAW+SPLINE (PS) \citet{Edelman:2021zkw}, which models the number density as:
\begin{equation}
     \frac{{\rm d} N}{{\rm d}\Theta} (\Lambda) \propto S(m_1,m_{min},\delta_m) PL(m_1,m_{min},m_{max},\alpha) e^{f(m_1|\{f_i\})},
\end{equation}
where $f(m_1|\{f_i\})$ is a cubic spline function and the $\{f_i\}$ are the values of this function at fixed knots, spaced log-uniformly. The number of knots was fixed to 20. Our PWP model differs from the PS by: (i) not assuming an underlying power-law shape, (ii) letting the number and the position of knots vary. Notice that both models are able to represent a simple power-law function.  

\begin{figure*}[t]
\centering
 \includegraphics[width=\textwidth]{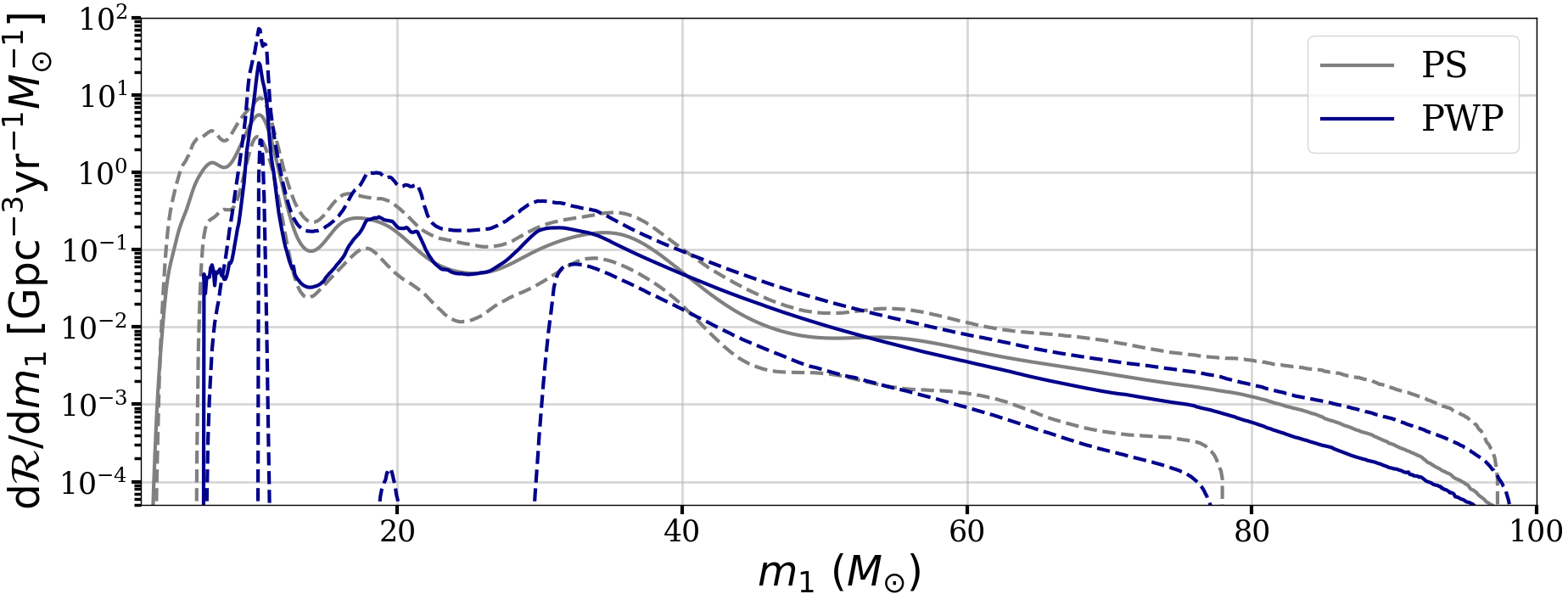}\\
 \centering
 \caption{Same as Fig.~\ref{fig:hba_lvk} for the comparison between the PS (grey) and PWP (blue) models. Both models agree on the location of the peak at low masses, with our model predicting a more peaked shape. They also agree on the location of the secondary peaks. The uncertainty in our model is larger, in particular around $20M_{\odot}$ due to less a priori asusmptions, e.g., we don't assume an underlying power-law shape.}\label{fig:hba_lvk_non_par}
\end{figure*}

\section{Inference on GWTC3}\label{sec:hba_gwtc3}
We apply both our PWP and FPG models to the LVK data. As described in Sec.~\ref{sec:hba}, for ease of comparison, we consider exactly the same events and the same samples as in the LVK analyses \citet{KAGRA:2021duu}. The inferred volumetric rates as a function of $m_1$ are shown in Fig.~\ref{fig:hba_lvk}, where they are compared to the fiducial LVK result. We recap how the volumetric rate is derived from the number density in App.~\ref{app:vol_rate}.   
\begin{figure}
\centering
 \includegraphics[width=0.5\textwidth]{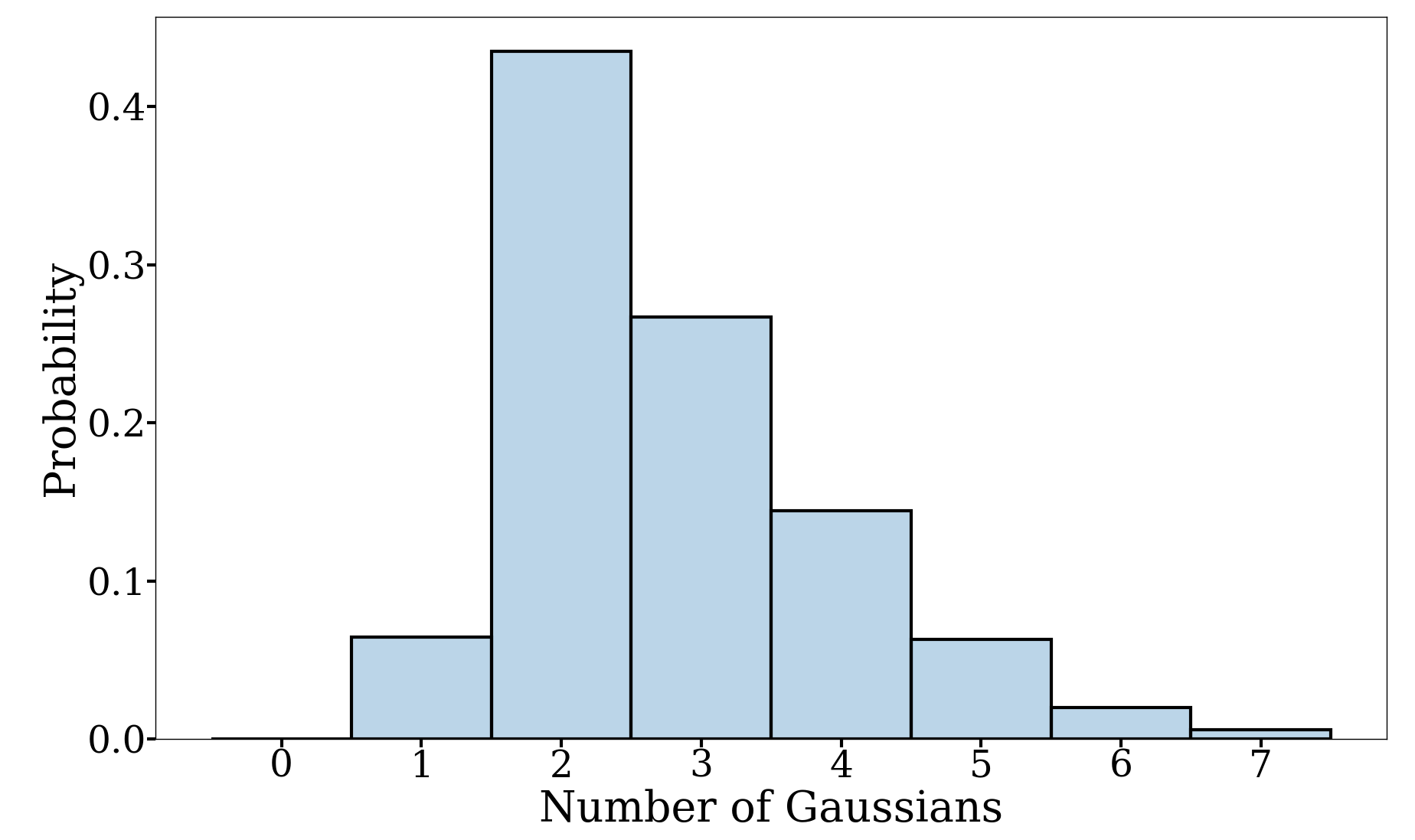}\\
 \centering
 \caption{Posterior distribution on the number of Gaussians for our FPG model. The model favours having 2 Gaussians: one at $\sim 35M_{\odot}$ and another at $\sim 10M_{\odot}$. It also has some support for a third Gaussian around $65M_{\odot}$. }\label{fig:hist_gauss_hba_lvk}
\end{figure}

\begin{figure}
\centering
 \includegraphics[width=0.5\textwidth]{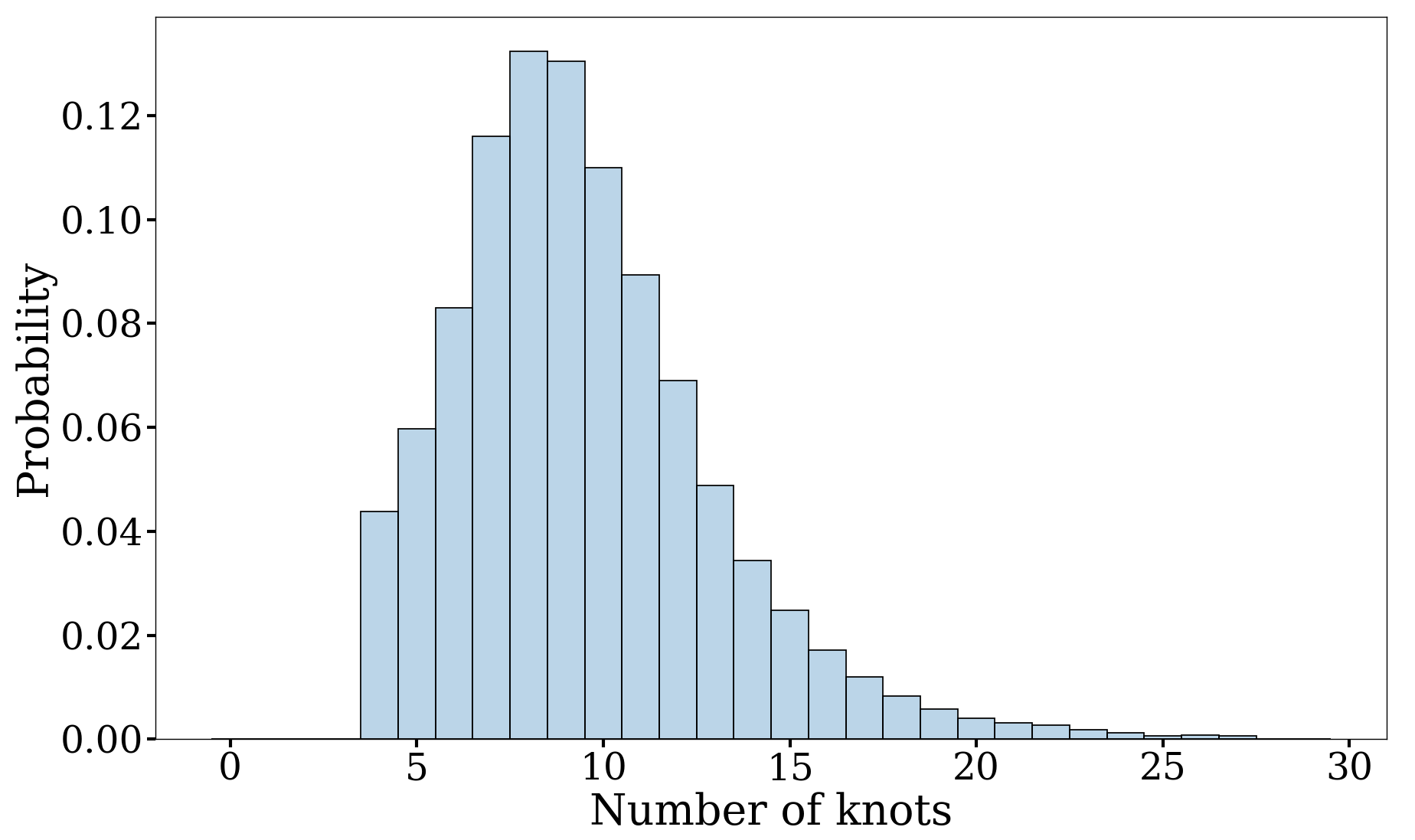}\\
 \centering
 \caption{Posterior distribution on the number of knots for our PWP model.}\label{fig:hist_knots_hba_lvk}
\end{figure}

The FPG model is in reasonable agreement with the LVK result, with the most noticeable difference at low masses. Regarding the power-law versus broken power-law comparison, the former is mildly favoured, with a Bayes’ factor of $1.671\pm 0.003$. However, in the broken power-law case, the posterior distribution is nearly flat in $m_{break}$ and is concentrated around $\alpha_1 = \alpha_2$, reducing to a simple power-law.  Thus, the favoured configuration is a power-law 
with 2 Gaussians, with a $(43.49 \pm 0.15)\%$ probability relative to all possible combinations. The second best is a power-law with 3 Gaussians, at $(23.91 \pm 0.11)\%$ probability. Finally, we find only a $(8.5 \pm 0.07)\%$ probability of having only Gaussians. 
In the preferred configuration, we find a Gaussian at $\sim35M_{\odot}$ and another at $\sim 10M_{\odot}$. As a consequence, the FPG model predicts a peak of events slightly displaced to larger masses compared to the fiducial LVK analysis. 
 As commented above and illustrated in Fig.~\ref{fig:hist_gauss_hba_lvk}, there is also some support for a third Gaussian around $65M_{\odot}$, leading to the observed excess at high masses. However, we note that the y-axis is in log-scale, which exaggerates the size of this effect. Finally, we do not find evidence for an excess around  $20M_{\odot}$. The result shown in Fig.~\ref{fig:hist_gauss_hba_lvk} is summed over all possible model component combinations, with relative probabilities obtained from sampling.

Our non-parametric model shows more differences with respect to the PP analysis. Its prediction resembles more the ones returned by the FLEXIBLE MIXTURES and POWER-LAW+SPLINE models of \citet{KAGRA:2021duu}. We compare the results of the latter to ours in Fig.~\ref{fig:hba_lvk_non_par} and in Fig.~\ref{fig:hist_knots_hba_lvk}, we show the posterior distribution on the number of knots for the PWP model. The PWP model agrees with the FPG and the PS models on the position of the main peak, but predicts a more pronounced shape. Such a peak is typically expected for binaries that form in isolation \citet{Giacobbo:2018etu,Wiktorowicz:2019dil,vanSon:2022myr}. Together, these results suggest that the PP model is not flexible enough to capture these fine features. The PWP also recovers a secondary maximum around $35 M_{\odot}$. It is slightly shifted to lower masses compared to the FPG, PP and PS models, but the $90\%$ confidence intervals still have a broad overlap in this region. Regarding the original goal of our analysis, the PWP model does suggest an excess of BHs around $20 M_{\odot}$, as indicated by the peak in the mean and the upper boundary of the $90\%$ confidence interval, in broad agreement with the PS prediction. We stress however that the uncertainty of the PWP model in this region is very large, and the inferred distribution is also compatible with not having a peak around $20M_{\odot}$. We attribute the larger uncertainty of our model with respect to the PS as due to the fact it makes fewer a priori assumptions, e.g., we do not assume that the underlying function is a power-law. Similarly, the difference in the volumetric rate before the peak is likely due to the underlying assumption on the shape of the $m_1$ pdf in the PS model. Given the small number of observations, this excess at $20M_{\odot}$ suggested by our model could be due to statistical fluctuations. In order to investigate this possibility, we perform a series of mock injections.

 \begin{figure*}
\centering
 \includegraphics[width=\textwidth]{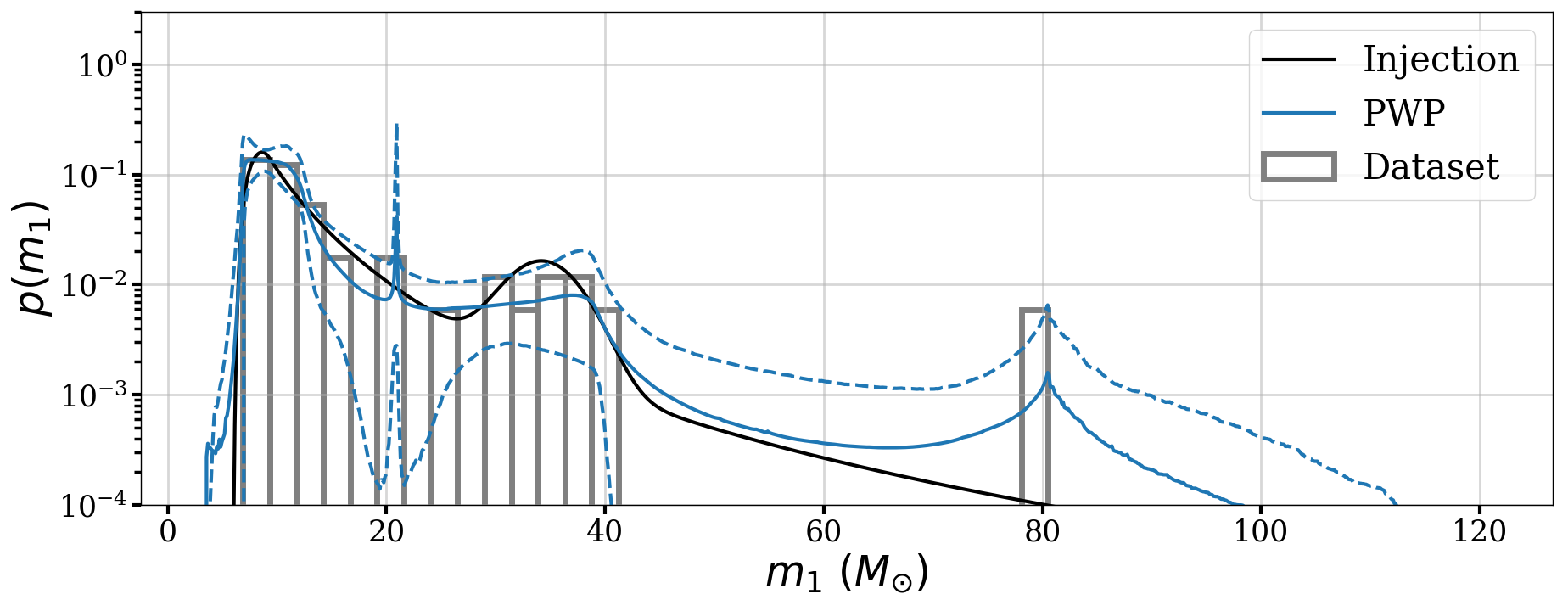}\\
 \centering
 \caption{Examples of a mock injection catalogue with 69 events where statistical fluctuations lead to erroneously thinking that there in an excess of BHs around $20 M_{\odot}$. The same happens around $80M_{\odot}$.}\label{fig:fake_peak}
\end{figure*}

\section{Inference on mock data}\label{sec:mock_hba}

Firstly, we want to assess how likely it is that we find a spurious peak-like feature between $13M_{\odot}$ and $25M_{\odot}$ in the distribution inferred with our PWP model when the underlying population does not have such a feature. This range is motivated by the results of \citet{Mahapatra:2022ngs} that show that BBHs with primary mass in this range likely contain at least one second generation BH. We decide on the presence of a peak-like feature by looking for a local maximum in the mean of the inferred pdf. We refer to \citet{Farah:2023vsc} for an analysis of the statistical significance of other noticeable features in the $m_1$ distribution. We draw 200 sets of hyperparameters from the LVK PP posterior and for each of those we simulate datasets of 69, 200 and 500 events. The latter two define a realistic range for the number of BBH observations we expect by O4. Then, we analyse the generated events with our non-parametric model and count the fraction of realisations in which we find a peak between $13M_{\odot}$ and $25M_{\odot}$ in the inferred distribution. Since we only want to get a rough idea of the significance of the peak observed in the LVK data, to speed up computations we do not account for selection effects nor measurement errors. We find that the probability to find a peak between $13M_{\odot}$ and $25M_{\odot}$ is: 
\begin{itemize}
    \item $0.04 \pm 0.01 $  with 69 events
    \item $0.02 \pm 0.01 $ with 200 events
    \item $<0.005$ with 500 events
\end{itemize}
We show in Fig.~\ref{fig:fake_peak} an example where applying our PWP model to a dataset of 69 events leads to erroneously thinking there is an excess of BHs around $20 M_{\odot}$. We show in gray the histogram of events. Although the underlying population has no local maximum around $20 M_{\odot}$, statistical fluctuations lead to an excess of events in this region, driving the erroneous inference. In this example, the same happens around $80M_{\odot}$.
Thus, the false-alarm probability for detecting an excess between $13M_{\odot}$ and $25M_{\odot}$ with our non-parametric model after observing 69 events is roughly $5\%$, but virtually null after 500 events. When analysing these same datasets with the FPG model, none of the cases shows a peak between $13M_{\odot}$ and $25M_{\odot}$. At high masses, where the events are more sparsely distributed, it is common for our non-parametric model to show spurious peaks, due to fitting for isolated events, as illustrated in Fig.~\ref{fig:fake_peak}. This problem does not concern the range we are interested in, between $13M_{\odot}$ and $25M_{\odot}$, which is in the bulk of distribution and where the events are more continuously distributed.

\begin{figure}
\centering
 \includegraphics[width=0.5\textwidth]{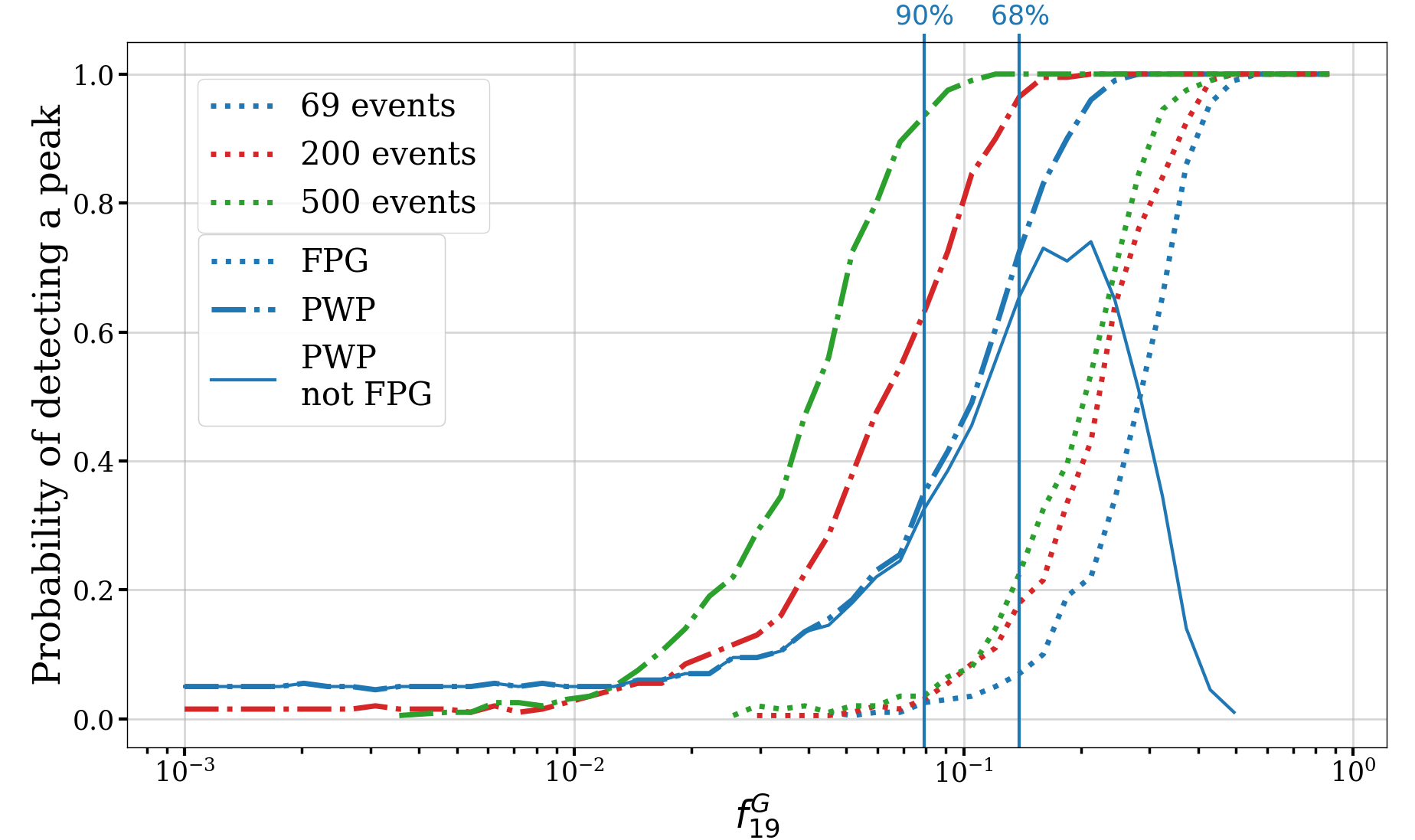}\\
 \centering
 \caption{Probability of detecting a peak between $13M_{\odot}$ and $25 M_{\odot}$ as a function of the weight of the Gaussian at $19 M_{\odot}$, for both the FPG and the PWP model, considering an increasing number of events. At small weights, the probability matches the false-alarm probability. The bumpy aspect at small weights is due to low statistics in this region, but the general trend is clear. Curves interrupted somewhere over the x-range go to 0. The solid blue curve shows the probability of detecting a peak with the non-parametric model and not with the semi-parametric one for 69 events, as it is the case for the GWTC-3 dataset. The vertical lines show the lower bounds on $f_{19}^G$ at $68\%$ and $90\%$ credibility, obtained by treating this curve as a likelihood, as described in the text.}\label{fig:probs_peak}
\end{figure}

Next, we want to estimate under which circumstances we can confidently detect the presence of the peak around $20 M_{\odot}$. For the same 200 samples from the LVK population, we define 50 new populations by adding a Gaussian of mean $19 M_{\odot}$ and width $2 M_{\odot}$ with increasing weight. For this purpose we define the normalised fraction $f_i=\lambda_i/\sum_i \lambda_i$, where the sum runs over the amplitudes of all components. In practice we grid the fraction $f_{19}^{G}$ logarithmically from $10^{-3}$ to 1 and redistribute the weight that was initially in the power-law between the Gaussian at $19 M_{\odot}$ and the power-law. The weight of the Gaussian at $\sim 35M_{\odot}$ remains the same. This simplistic procedure is driven by the idea that if there is an additional peak around $20 M_{\odot}$, it has been "swallowed" by the power-law component. One might also expect that the inferred power-law index is actually smaller than it should, i.e., the power-law is less steep in order to accommodate the excess. However, this effect should be small and we are looking for a rough estimate of the detectability of an excess around $20 M_{\odot}$, so we do not account for it. Once again we draw sets of 69, 200 and 500 events from each of these populations, and analyse them both with the PWP and with the FPG model. Fig.~\ref{fig:probs_peak} shows the probability of finding a peak between $13M_{\odot}$ and $25M_{\odot}$ as a function of the weight of the Gaussian around $19 M_{\odot}$ with each model. At small weights, we recover the false-alarm probability discussed above. As expected, the chances of detecting an excess increase as the weight increases. With 500 events, our non-parametric model will reliably indicate if the excess is truly physical, with higher probability than the FPG model. From this plot, we also see that it is not surprising that the latter does not find the putative excess of BHs in the current LVK data, since it requires large weights to identify it with only 69 events. On this same figure, the solid blue curve shows the probability of detecting a peak with the non-parametric model and not with the semi-parametric one when observing 69 events, as it the case for the GWTC-3 dataset. Under the hypothesis that the BBH population is compatible with the ones used to perform the mock injections, this can be interpreted as a likelihood for the observed datum that one fit finds a peak and the other does not. Constructing a posterior from this likelihood and a flat prior on $f_{19}^{G}$, we estimate that $f_{19}^{G} \geq 0.08$ at $90\%$ credibility and $f_{19}^{G} \geq 0.14$ at $68\%$ credibility. These are the values indicated with vertical lines. From the value at which the full line curve goes to 0, we deduce in addition that $f_{19}^{G} < 0.5$.

Compared to the full analysis of Sec.~\ref{sec:hba_gwtc3}, our mock studies do not take into account the presence of selection effects and measurement uncertainties in individual events. We expect the latter to smooth out the mass distribution, making fine features harder to detect. Thus, we might be slightly overestimating the detectability of peaks by our methods. Measurement uncertainties would also smooth out sharp features like the one seen in Fig.~\ref{fig:fake_peak}. Selection effects would make the distribution of observed events different to the astrophysical one. For instance, events in the peak around $10 M_{\odot}$ would be harder to detect compared to events with mass $20M_{\odot}$ or $35M_{\odot}$. On average, we could expect this to decouple from the statistical fluctutations we investigated in the first part of this section, and not have too much impact on our conclusions regarding the detectability of spurious peaks. Quantities like the position of the peaks and the "excess probability" in these peaks would in turn be affected, but these are not the focus of our investigation. On the other hand, selection effects would make it easier to identify real excesses around $20M_{\odot}$ for a fixed number of observed events.


\section{Conclusion}\label{sec:ccl}

 Capturing distinctive features in the population of BBHs is crucial to discriminate between astrophysical models. Astrophysically motivated priors are useful to directly constrain physical parameters, but lack flexibility when the population includes unanticipated structure. In this paper we have proposed two flexible models for the distribution of primary masses and applied them to GWTC-3. A crucial ingredient of our method is RJMCMC, which allows the complexity of the model to be chosen by data as well as to perform model selection. With these models in hand, we have assessed the statistical significance of the excess of events around $15-20 M_{\odot}$ found by some of the population analyses performed on GWTC-3 \citet{KAGRA:2021duu,Edelman:2022ydv,Tiwari:2020otp}, but not all \citet{Sadiq:2021fin,Ruhe:2022ddi,Callister:2023tgi}. 
 
 Our first model is an extended and more flexible version of the POWER-LAW+PEAK model of the LVK, where the number of Gaussians is free to vary and we can choose between having a power-law, a broken power-law or none. It illustrates how RJMCMC can be used to perform model selection, allowing to choose between a variety of models at once, instead of running them individually and comparing the evidences afterwards. We have found that the current data favours having a power-law component with two Gaussians: one around $35M_{\odot}$ and another around $10M_{\odot}$. As a consequence, it predicts a peak of events at slighlty higher masses than in the fiducial LVK analysis. There is also mild support for a third Gaussian at $\sim 65M_{\odot}$, but it has small significance in the current data. Furthermore, it is disfavoured to have only Gaussians. Finally, we find no sign for a break in the slope of the power-law, but this might change as the number of observations increases. Moreover, more elaborated combinations of parametric functions could be considered in order to fully take advantage of the flexibility offered by RJMCMC.
 
 Next, we have proposed a non-parametric model representing the $m_1$ pdf as a piece-wise power-law function. We infer the position of the knots and the value of the pdf at those knots, but also the number of knots, thanks to the RJMCMC. The complexity of our model is not pre-determined, but it is decided by data. This model shows a few differences with respect to the POWER-LAW+PEAK model, in particular it agrees with our semi-parametric model regarding the displacement of the low-mass peak and also suggests an excess of BHs around $20 M_{\odot}$. It is in better agreement with the POWER-LAW+SPLINE model of the LVK. However, by performing mock-injections under simplifying assumptions (i.e. negelcting selection effects and measurement uncertainties), we have found that there is roughly a $5\%$ chance that the peak at $20 M_{\odot}$ is due to statistical fluctuations when assuming a population compatible with the LVK POWER-LAW+PEAK analysis. With 500 events the false-alarm probability is nearly zero. Moreover, when analysing mock-populations that do have an excess at $20M_{\odot}$, our model can more easily find it than our FLEXIBLE POWER-LAW+GAUSSIANS model. As the number of events increases, additional features in the $m_1$ distribution might appear. Hierarchical mergers could lead to a series of regularly spaced peaks with decreasing amplitude \citet{Tiwari:2020otp,Tiwari:2021yvr}. Detecting more massive events such as GW190521 \citet{LIGOScientific:2020iuh} will inform us on the mass-gap, and whether there is a dearth of events in this region, as predicted by models. Our non-parametric model would be able to capture these features without any a priori modelling of the signature of these effects on the $m_1$ distribution, proving a powerful tool to better understand astrophysics. 

In this work we have assumed the population prior to be separable in the event parameters, primary mass, mass ratio, spins and redshift. However, finding correlations between parameters would increase our ability to discriminate between astrophysical scenarios. Different approaches have already been explored \citet{Hoy:2021rfv,Callister:2021fpo,Fishbach:2021yvy,Adamcewicz:2022hce,Bavera:2022mef,Biscoveanu:2022qac,Ray:2023upk}, with some of them yielding first hints of correlations. With the increase in the number of events, such fine features will become more prominent. It is therefore the natural next step for us to extend out method to multi-dimensional distributions.

 \section*{Acknowledgments}

This research has made use of data or software obtained from the Gravitational Wave Open Science Center (gwosc.org), a service of LIGO Laboratory, the LIGO Scientific Collaboration, the Virgo Collaboration, and KAGRA. LIGO Laboratory and Advanced LIGO are funded by the United States National Science Foundation (NSF) as well as the Science and Technology Facilities Council (STFC) of the United Kingdom, the Max-Planck-Society (MPS), and the State of Niedersachsen/Germany for support of the construction of Advanced LIGO and construction and operation of the GEO600 detector. Additional support for Advanced LIGO was provided by the Australian Research Council. Virgo is funded, through the European Gravitational Observatory (EGO), by the French Centre National de Recherche Scientifique (CNRS), the Italian Istituto Nazionale di Fisica Nucleare (INFN) and the Dutch Nikhef, with contributions by institutions from Belgium, Germany, Greece, Hungary, Ireland, Japan, Monaco, Poland, Portugal, Spain. KAGRA is supported by Ministry of Education, Culture, Sports, Science and Technology (MEXT), Japan Society for the Promotion of Science (JSPS) in Japan; National Research Foundation (NRF) and Ministry of Science and ICT (MSIT) in Korea; Academia Sinica (AS) and National Science and Technology Council (NSTC) in Taiwan.

\section*{Data Availability}

This work makes use of the publicly available data released by the LVK \citet{data_gwtc3,data_gwtc3_pop} and the publicly available code Eryn \citet{eryn}. In addition to our own, we show results from \citet{KAGRA:2021duu}, available on \citet{data_gwtc3_pop}.


\bibliographystyle{mnras}
\bibliography{Ref.bib} 

\appendix

 \clearpage
 
\section{Population priors}\label{app:pop_priors}

We describe the models used for the single event parameters other than the primary mass, these are the same as in the fiducial analysis of \citet{KAGRA:2021duu}. For the mass ratio, we assume a power-law distribution:
\begin{equation}
    p(q|m_1,\Lambda_q) \propto q^{\beta}S(qm_1,m_{min},\delta_m)
\end{equation}
with
\begin{equation}
S(m_1,m_{min},\delta_m) =
	\begin{cases}
		& 0 ;  \ {\rm if} \; m_1 < m_{min}  \\
  & [f(m_1-m_{min},\delta_m)+1]^{-1}  \\
  &\ {\rm if} \;  m_{min} \leq m_1 \leq m_{min}+\delta_m  ;  \\ 
		& 1 ; \ {\rm if} \; m_1 > m_{min}+\delta_m , \label{eq:smooth}
	\end{cases}
\end{equation} 
and
\begin{equation}
    f(m,\delta_m)=e^{\frac{\delta_m}{m}+\frac{\delta_m}{m-\delta_m}}.
\end{equation}

In the PWP model, we do not use the smoothing function, and impose a sharp cut-off: 
\begin{equation}
   p(q|m_1,\Lambda_q) 
	\begin{cases}
             & 0 ; \ {\rm if} \; qm_1 < 2 \\ 
		& \propto q^{\beta} ;  \ {\rm otherwise.}  \\
	\end{cases}
\end{equation}

We assume a common Beta-dsitribution for the spins magnitude:
\begin{equation}
    p(\chi|\Lambda_{\chi})={\rm Beta}(\alpha_{\chi},\beta_{\chi}).
\end{equation}
As for their orientation, we model the joined distribution of the cosines of the tilt angles as a mixture between a two-dimensional Gaussian of width $\sigma_t$, centred at 0 and truncated at -1 and 1, and a two-dimensional flat distribution between -1 and 1.:
\begin{align}
    p(\cos(\theta_1),\cos(\theta_2)|\Lambda_{\theta})=&\zeta G_t(\cos(\theta_1),\cos(\theta_2),0,\sigma_t) \nonumber \\
    &+(1-\zeta)U(-1,1)
\end{align}
Finally, we assume that the rate of BBHs evolves with redshift as $\mathcal{R}(z)=(1+z)^{\kappa}$, which leads to the pdf on $z$:
\begin{equation}
    p(z|\Lambda_z) \propto (1+z)^{\kappa-1} \frac{{\rm d} V_c}{{\rm d}z},
\end{equation}
where $V_c$ is the comoving volume. 

The priors for $\beta$, $m_{min}$, $\delta_m$, $\zeta$, $\sigma_t$ and $\kappa$ are taken to be flat. For $\alpha_{\chi}$ and $\beta_{\chi}$, we assume the prior on the mean $\mu_{\chi}$ and the variance $\sigma_{\chi}^2$ of the Beta-distribution to be flat, subject to the condition $\alpha_{\chi}>1$, $\beta_{\chi}>1$.

\section{Volumetric rate}\label{app:vol_rate}

The number density can be transformed into a volumetric rate by taking its ratio with the observed space-time volume $VT_{\rm obs}$, defined as:
\begin{equation}
    VT_{\rm obs}=T_{\rm obs} \int_{0}^{z_{\rm max}} \frac{1}{1+z} \frac{{\rm d}V_c}{d_z} \mathcal{R}(z) \ {\rm d} z
\end{equation}
We take $z_{\rm max}=2.3$ and use the value of $T_{\rm obs}$ provided in the public data released along \citet{KAGRA:2021duu}.
Finally,
\begin{equation}
    \frac{{\rm d}\mathcal{R}}{{\rm d}m_1}(\Lambda)= \frac{1}{VT_{\rm obs}}\frac{{\rm d}N}{{\rm d}m_1}(\Lambda).
\end{equation}

\bsp	
\label{lastpage}
\end{document}